\documentclass[11pt,prerint]{elsarticle}
\usepackage{epsfig}
\usepackage{bm}
\usepackage{amssymb}
\usepackage{wrapfig}
\usepackage{ulem}
\usepackage{multirow}

\newcommand{\el}{{\cal L}}

\newcommand{\psibar}{\mbox{$\overline{\psi}$}}

\newcommand{\epsi}{\mbox{$\varepsilon$}}

\newcommand{\vp}{\mbox{$\bm{p}$}}

\newcommand{\vk}{\mbox{$\bm{k}$}}

\newcommand{\vB}{\mbox{$\bm{B}$}}

\newcommand{\Kp}{$K^+$}

\journal{Physics Letters B}

\begin{document}

\begin{frontmatter}

\title{\bf Asymmetric Neutrino Production \\
in Strongly Magnetized Proto-Neutron Stars}

\author[nubs,nao]{Tomoyuki~Maruyama}

\author[nao]{Jun~Hidaka\fnref{fn1}}

\author[nao,asTky]{Toshitaka~Kajino}

\author[CIT]{Nobutoshi~Yasutake}

\author[nao]{Takami~Kuroda\fnref{fn2}}

\author[snuv]{Myung-Ki Cheoun}

\author[hyuv]{Chung-Yeol Ryu}

\author[ND,nao]{Grant J. Mathews}

\date{\today}

%\pacs{25.30.Pt, 26.60.-c, 24.10.Jv}

\begin{abstract}
We calculate the neutrino production cross-section in
 proto-neutron star matter in the presence of a strong magnetic field.   
We assume isoentropic conditions and introduce a new equation of 
state parameter-set in the relativistic mean-field approach 
that can reproduce neutron stars with $M >  1.96$ M$_\odot$ as required by observations. 
We find that the production process  increases the flux of emitted neutrinos
along the direction parallel to the magnetic field and decreases
 the flux in the  opposite direction.
This  means that the neutrino flux asymmetry due to the neutrino absorption 
and scattering processes  in a  magnetic field becomes larger 
by the inclusion  of the neutrino production process.
\end{abstract}

\begin{keyword}
neutrino production \sep strong magnetic field \sep relativistic mean field approach
\end{keyword}

\address[nubs]{College of Bioresource Sciences,
Nihon University,
Fujisawa 252-8510, Japan}
\address[nao]{National Astronomical Observatory of Japan, 2-21-1 Osawa, Mitaka, Tokyo 181-8588, Japan}
\address[asTky]{Department of Astronomy, Graduate School of Science, University of Tokyo, Hongo 7-3-1, Bunkyo-ku, Tokyo 113-0033, Japan
}

\address[CIT]{ Department of Physics, Chiba Institute of Technology, 
2-1-1 Shibazono, Narashino, Chiba 275-0023, Japan}
\address[snuv]{Department of Physics, Soongsil University, Seoul, 156-743, Korea}
\address[hyuv]{General Education Curriculum Center, Hanyang University, Seoul, 133-791, Korea}
\address[ND]{Center of Astrophysics, Department of Physics,
University of Notre Dame, Notre Dame, IN 46556, USA}

\fntext[fn1]{Present Address: Meisei University, Hino, Tokyo 191-8506, Japan}
\fntext[fn2]{Present Address: Department of Physics, University of Basel, CH-4056 Basel, Switzerland}

\end{frontmatter}
%\maketitle

%\tableofcontents

%{\bf INTRODUCTION - yet to be rewritten}

The magnetic field in neutron stars plays an important role in the interpretation of many observed phenomena.  
Indeed, strongly magnetized neutron stars (dubbed  {\it magnetars}
\cite{mag1,pac92,mag3}) hold the key to understanding the asymmetry in
supernova (SN) remnants, pulsar kicks~\cite{lyne94}, and the  still
unresolved mechanism of non-spherical SN explosions. 
Such strong magnetic fields are also closely related to the  unknown
origin of the kick velocity~\cite{rothchild94} that proto-neutron stars
(PNSs) receive at  birth.
Although several post-collapse instabilities have been studied
as a possible source to trigger a non-spherical explosion
leading eventually to a pulsar kick,
there remain  uncertainties in the  global initial asymmetric perturbations
and the numerical simulations
~\cite{burrows06,marek09}.

In strongly  magnetized PNSs,
asymmetric neutrino emission emerges from parity violation
in the weak interaction~\cite{vilenkin95,horowitz98}
and/or an asymmetric distribution of the magnetic field~\cite{bisnovat93}.
Indeed, recent theoretical calculations~\cite{arras99,lai98} have suggested
that even a $\sim$1\% asymmetry in the neutrino emission out of total neutrino
luminosity $\sim 10^{53}$ ergs might be enough
to explain the observed pulsar kick velocities.

In our previous work \cite{MKYCR11,MYCHKMR12},
we calculated neutrino scattering and absorption cross sections in hot, 
dense magnetized neutron-star matter including hyperons
in a fully relativistic mean field (RMF) theory~\cite{serot97}.
We evaluated both the associated pulsar kick velocities \cite{MYCHKMR12} 
and the spin deceleration \cite{MHKYKCRM13} for PNSs.
The magnetic field was shown to enhance the scattering cross-section
in a direction parallel to the magnetic field for the final neutrino momentum
and to reduce the absorption cross-section
along  the same direction for the initial neutrino momentum.
When the neutrino momentum  is anti-parallel to the magnetic field, the  opposite effect occurs.
For  a magnetic field strength of  $B = 2 \times 10^{17}$G and densities in excess of nuclear matter 
$\rho_B \approx 1 -5 \rho_0$, the enhancement in the scattering
cross-section was calculated  to be  about 1\%  \cite{MYCHKMR12},
while  the reduction in  the neutrino absorption was 2 $-$ 4\%. 
This enhancement and reduction were conjectured to 
increases the neutrino momentum flux emitted along the north magnetic pole, 
while  decreasing the flux along the south pole  
when the magnetic field has a poloidal distribution.
By exploiting a one-dimensional  Boltzmann equation in the 
attenuation approximation and including only neutrino absorption,
we estimated that the pulsar kick velocity is about 520 km/s for a star 
with baryon mass $M_{NS} = 1.68$ M$_{\odot}$, $B = 2 \times 10^{17}$ G, 
$T=20$ MeV, and $E_T \approx 3 \times 10^{53}$erg.

In those calculations, however, we did not consider the neutrino production 
process through the direct URCA (DU) and modified URCA (MU) processes.
A strong magnetic field may lead to an angular-dependence of
the neutrino production in the URCA process
because of the spin polarization of electrons and positrons in matter
\cite{chugai84,dorofeev}.
It has also been reported \cite{kisslinger2,kisslinger3}  that the Landau levels due to a magnetic field
can cause  an asymmetry in the 
neutrino emission which causes a pulsar kick velocity.
Furthermore, an angular dependence of the neutrino production caused by a magnetic field
has even been  reported \cite{vosk86,parenkov} to occur 
in a  pion condensation phase  or 
in a  quark-matter color-super conducting phase \cite{berderman}, {\it etc}.

Therefore, the neutrino production process in the presence of a  magnetic field may also lead to  asymmetric neutrino emission from PNSs. 
In this report, we take this production process into account in our model by
calculating the cross-sections using
the PM1L1 parameter-set \cite{kpcon} with an isothermal neutron-star model.
However, including $\Lambda$ particles in this parameter-set cannot  
reproduce the observed  neutron star mass of 1.97 M$_\odot$
 for PSR J1614-2230 \cite{T-Solar}.
In addition, an isothermal model gives a temperature that is too high in
the low density region. 
In this work, therefore, we first improve the RMF parameter-set 
to allow a more massive neutron star.  
We then study the neutrino absorption and production though the DU process 
using a relativistic mean-field (RMF) approach in an isoentropic neutron star.

We start from the RMF Lagrangian comprised of  nucleons,
$\Lambda$ fields, sigma and omega meson fields, and the iso-scalar and
Lorentz vector interaction between nucleons.
We parameterize  the nucleon mean-fields
to reproduce the consensus  nuclear-matter  properties, i.e.
a binding energy of 16 MeV, a nucleon effective mass of $M^*/M = 0.65$,
and a  incompressibility coefficient of  $K=250$ MeV  in symmetric 
nuclear matter  at a saturation density of $\rho_0 = 0.17$fm$^{-3}$.
In  the analysis of heavy-ion experiments \cite{TOMO1}
an EOS with $M^*/M =0.65$ and $K=200 -400$MeV simultaneously reproduces the results of
the transverse-flow and sub-threshold \Kp-production experiments.
In the analysis of such data one needs to consider 
the momentum-dependence of the RMF approach.  
However, such momentum dependence only affects observables at an energy 
above a few hundred MeV per nucleon and 
does not significantly  affect the properties of infinite nuclear matter at the temperatures experienced by
proto-neutron stars.

The $\sigma-\Lambda$ and $\omega-\Lambda$
couplings are taken to be 2/3 that of the  nucleon, {\it i.e.}
$g_{\sigma, \omega}^\Lambda = 2/3 g_{\sigma, \omega}$
by taking account of the quark degrees of freedom.
For the $\Lambda$-$\Lambda$ interaction we use
$h_{s}=0.3467 g_{\sigma}$ and $h_{v} = 0.5 g_{\omega}$.

In order to stiffen the  EOS when  including the lambda ($\Lambda$) particles,
we introduce an additional $\Lambda$-$\Lambda$ interaction term  with a 
  Lagrangian  written as
\begin{equation}
\el_{\Lambda \Lambda} =  \frac{h_{s}^2}{2m_s^2}
\left\{\psibar\Lambda \psi_\Lambda \right\}^2
+   \frac{h_{v}^2}{2m_v^2}
\left\{\psibar\Lambda \gamma_\mu \psi_\Lambda \right\}
\left\{\psibar\Lambda \gamma^\mu \psi_\Lambda \right\},
\label{LLint}
\end{equation}
where $h_{s}$ and $h_{v}$ are the scalar and vector couplings 
between the two $\Lambda$s, respectively. 
The scalar and vector meson masses are taken to be
$m_s = 550$MeV and $m_v = 783$MeV, as in  previous
calculations \cite{MKYCR11,MYCHKMR12,MHKYKCRM13}.
Of course the new EOS reproduces massive neutron stars with 2.0 solar masses
even if the nuclear medium is composed of $\Lambda$ particles.

\begin{figure}[ht]
%\vspace{0.5cm}
%\hspace*{0.5cm}
\begin{center}
{\includegraphics[scale=0.5,angle=270]{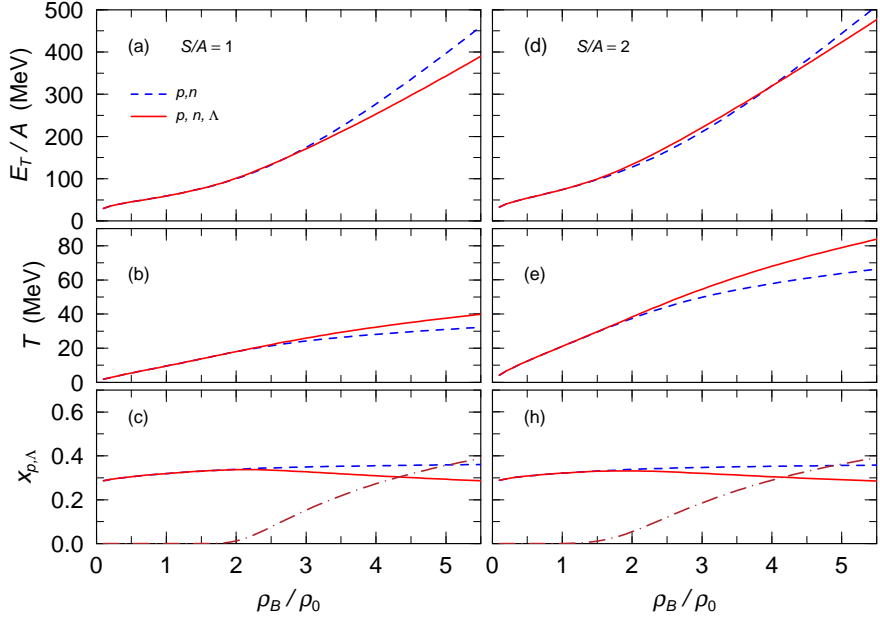}}
\caption{\small (Color online)
Upper panels (a) and (d):  Density dependence of the total energy per baryon
$E_T/A$ of  neutron-star matter for entropies $S/A$ = 1 (a) and 2 (d).
Middle panels (b) and (e): Temperature profiles for \ entropies $S/A$ = 1 (b)
and 2 (e).
Solid and dashed lines represent results with and without $\Lambda$ particles, 
respectively, in the EOS.
Lower panels (c) and (f):  Number fractions of
protons $x_p$ and of $\Lambda$ particles $x_\Lambda$
for entropies $S/A$ = 1 (c) and 2 (f).
Solid and dashed lines stand for $x_p$ with and without $\Lambda$ particles, 
respectively. 
Dot-dashed lines show the $x_{\Lambda}$ fraction.}
\label{EOSf}
%Fig.1
\end{center}
\end{figure}

Figure~\ref{EOSf} shows total energies per baryon $E_T/A$, 
temperature profiles, and number fractions for  various constituent particles 
in an  isoentropic system 
with entropy per baryon $S/A$ = 1 or 2.
The  proton fraction is $x_p \approx 0.3$ in all density region.
When one includes $\Lambda$s in the system,  they appear at a density $\rho_B\ ^>_\sim2 \rho_0$
and the number  fraction $x_\Lambda$ increases with increasing density.
In these isoentropical models, the proton fraction slightly decreases even when the $\Lambda$s appear, while in an isothermal model
 $x_p$ decreases more rapidly.

Using the above EOS we calculated the neutrino absorption and
production cross-sections.
In this work we assume a uniform dipole magnetic field along the
$z$-direction, i.e.~$\vB = B {\hat z}$. Since even for an astronomically 
strong magnetic field the associated energy scale is still much weaker than 
the strong interactions, $\sqrt{e B} \ll \mu_a$, where $\mu_a$ is the
chemical potential of the particle $a$,
we can treat the magnetic field perturbatively. 
Hence, we ignore the contribution from the convection current and
only consider the spin-interaction.
The wave function for a baryon $b$
in a strong magnetic field is obtained by solving the following Dirac equation
\begin{equation}
\left[ \gamma_\mu p^\mu - M^*_b -  U_0 (b) \gamma_0 - \mu_b B \sigma_z \right ] u_b(p,s) =0 ,
\label{DiracE}
\end{equation}
with $M^*_b = M_b - U_s (b)$.
The quantities, $U_s(b)$ and $U_0(b)$, are the scalar mean-field and
the time-component of the vector mean-field, respectively.
These scalar and vector fields are calculated within RMF theory. 
We set $B = 10^{17}$G as a representative maximum field strength
inside a neutron star.
This value corresponds to $\mu_N B = 0.32$ MeV which satisfies 
$|\mu_b B| \ll \epsi_{\nu} \ll E_b^*(\vp) \equiv \sqrt{\vp^2 + M_b^{*2}}$.
The initial momentum here is taken to be the chemical
potential $|\vk_i| = \epsi_{\nu}$.

We calculated the absorption and scattering neutrino cross-sections
perturbatively, and separated  the cross section into the two parts:
%\begin{equation}
$\sigma_{S,A} =  \sigma^0_{S,A} + \Delta \sigma_{S,A}$,
%\end{equation}
where  $\sigma^0_{S,A}$ is independent of $B$, and
$\Delta \sigma_{S,A}$ is proportional to $B$. 
The subscripts $S$ and $A$ indicate  scattering ($\nu_e \rightarrow \nu_e$)
or absorption ($\nu_e \rightarrow e^{-}$), in obvious notation.
Related weak couplings are taken from Ref.~\cite{rml98}.

%\begin{figure}[t]
%\begin{wrapfigure}{r}{8.6cm}
\begin{figure}[ht]
\begin{center}
\includegraphics[scale=0.5,angle=270]{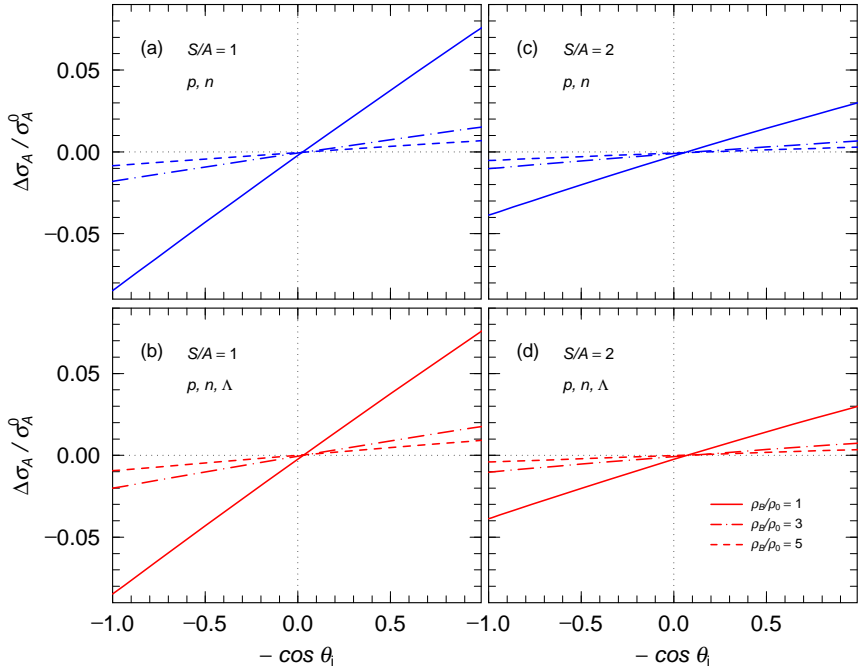}
\caption{\small{(Color online)
 Ratio of the magnetic part of the absorption cross-section
 $\Delta \sigma_A$ to the cross-section without
a magnetic-field $\sigma_A^0$. Lines are drawn for matter without $\Lambda$s (a) and
with $\Lambda$s (b) at  $T=20$MeV.
Solid, dot-dashed and dashed lines represent the results at
$\rho_B = \rho_0$, $3 \rho_0$ and $5 \rho_0$, respectively.
Neutrino incident energies are taken  to be equal to the neutrino chemical
potentials. }
}
\label{CrMag}
\end{center}
%Fig.2
\end{figure}
%\end{wrapfigure}

In Fig.~\ref{CrMag}, we show the magnetic part of the absorption cross-section
(a and b) as a function of the initial neutrino angle
for entropy $S/A = 1$  without  (a) and with  (b) $\Lambda$ particles in the EOS,
and for $S/A = 2$ without (c) and with (d) $\Lambda$ particles.

At  $\rho_B=\rho_0$, the magnetic field suppresses the absorption cross-sections in a direction parallel 
to the magnetic field ${\vB}$ by  about 8.3\% for an entropy of $S/A = 1$ and by about 4 \% for $S/A =2$. 
Hence, the magnetic field increases the emitted 
neutrino flux  in the direction of the north magnetic pole and decreases the flux along the south magnetic pole.
The suppression of $\sigma_A$ for $S/A=1$ at $\rho_B = \rho_0$
turns out to be much larger than in an isothermal model with $T=20$ MeV, 
because the temperature at nuclear matter density in the isoentropic model  is 
only about $T=7$MeV. 
At lower temperature  the magnetic contribution becomes larger. 
However, at higher densities and temperatures the suppression is comparable  
in the two models.

As discussed in Ref.~\cite{MYCHKMR12},
the normal parts of the cross-sections, $\sigma_0$, decrease
as the temperature and the density become lower.
In contrast, the magnetic parts, $\Delta \sigma$, increase
as the temperature becomes lower.  Also, as the density decreases the magnetic part decreases more slowly than the
normal part.  This is
because $\Delta \sigma$ is
approximately proportional to the fractional area of the distorted
Fermi surface caused by the magnetic field.  
Because of these two effects, the relative strength 
$\Delta \sigma_A/\sigma^0_A$ 
becomes significantly larger when the density and entropy are small.

\bigskip

Now considering  the neutrino production process,
we define the integrated cross-section for neutrino production as follows,
\begin{equation}
\sigma_{pr} (\vk_\nu)
= \int \frac{d^3 k_i}{(2 \pi)^3} n_e(e_i(\vk_i))
\frac{d^3}{d \vk_\nu^3} \sigma_{pr} (\vk_\nu, \vk_i) ,
\label{ph-nt-prod}
\end{equation}
where $e_i (\vk_i)$ is the single particle energy of electrons with
momentum $\vk_i$. The cross-section and the electron momentum
distribution function in the presence of a  magnetic field are separated 
in a perturbative way into the two parts
\begin{eqnarray}
\frac{d^3 \sigma_{pr}}{d k_{\nu}^3} &=& \frac{d^3 \sigma_{pr}^0}{d k_{\nu}^3}
+ \frac{d^3 \Delta \sigma_{pr}}{d k_{\nu}^3} ,
\\
n_{e} (e_i(\vk_i)) &=& n_{e} (|\vk_i|)+ \Delta n_e(\vk_i)~ .
\end{eqnarray}
The first terms are independent of the $B$ field, and the second terms are
proportional to $B$. Then, the neutrino phase-space distribution for the DU process in the presence of a  magnetic field
 also separates into the two parts 
\begin{eqnarray}
\sigma_{pr} (\vk_\nu) &\approx& \sigma^0_{pr} (\vk_\nu)
+ \Delta  \sigma_{pr} (\vk_\nu)
\nonumber \\
&\approx& \int \frac{d^3 k}{(2 \pi)^3}
n_e^{(0)}(\vk) \frac{d^3 \sigma^0_{pr}}{d \vk_\nu^3 }
+ \int \frac{d^3 k}{(2 \pi)^3} \left[
n_e^{(0)}(\vk) \frac{d^3 \Delta \sigma_{pr}}{d \vk_\nu^3 }
+ \Delta n_e(\vk) \frac{d^3 \Delta \sigma^{(0)}_{pr}}{d \vk_\nu^3 }
\right] .
\label{ph-nt-pdM}
\end{eqnarray}

Detailed expressions for $e_i$ and $\Delta n_e$ are
 given in Ref.~\cite{MHKYKCRM13}.
As shown in Ref.~\cite{MHKYKCRM13}, we can obtain  the cross-section for $e^{-} + B_i \rightarrow B_f + \nu_e$  by
exchanging the lepton chemical potentials for the neutrinos and electrons
in the cross-section  for  $\nu_e + B_i \rightarrow B_f + e^{-}$.

\begin{figure}[ht]
\vspace{0.5cm}
%\hspace*{0.5cm}
\begin{center}
{\includegraphics[scale=0.5,angle=270]{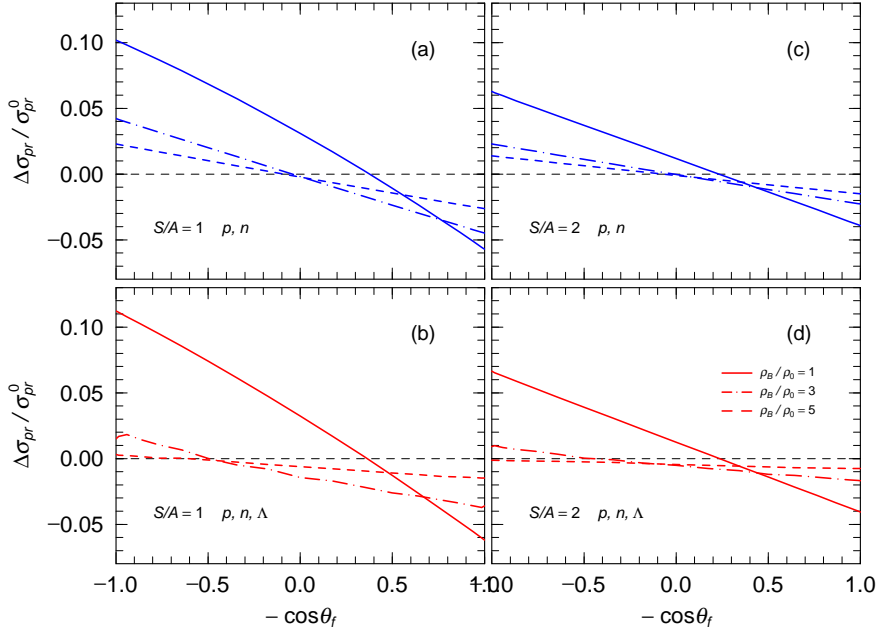}}
\caption{\small
Normalized magnetic part of the total production cross section ,
$ \Delta \sigma_{pr} / \sigma_{pr}^0$,
as a function of final neutrino angle $\theta_{f}$
in a system without hyperons for the entropy $S/A=1$ (a) and $S/A=2$ (b), and
with hyperons at $S/A=1$ (c) and $S/A=2$ (d).
In each panel the solid, dash-dotted and dashed lines
represent the results for  $\rho_B = \rho_0$, $3 \rho_0$ and $5 \rho_0$,
respectively.
A magnetic field strength of $B=10^{17}$G was used in this calculation.}
\label{PdMag}
%Fig.3
\end{center}
\end{figure}

Fig.~\ref{PdMag} shows the magnetic part of the integrated production
cross-section
$\Delta \sigma_{pr}$ normalized to $\sigma^0_{pr}$ as a function of
$\theta_f$.  Calculations were made for densities in the range 
$\rho_0 \le \rho_B \le 5 \rho_0$ as indicated.
Final neutrino energies were  taken to be equal to the neutrino chemical
potential, $|\vk_\nu| = \epsi_\nu$.

At $\rho_B = \rho_0$ with  entropy $S/A = 1$, the magnetic part is enhanced 
by about 10 \% for  $\theta_\nu = 0^\circ$ 
and suppressed by about 6 \% for  $\theta_\nu = 180^\circ$.  
This is true for both systems with and without $\Lambda$ particles.
Hence,  the magnetic-field gives rise to an about 8 \% asymmetry 
in the production process.
As the density increases, the magnetic contribution becomes smaller, 
particularly in the system with  $\Lambda$ particles.
For $S/A=2$, the asymmetry is about 6 \% at $\theta_\nu =0$ 
and 4 \% for $\theta_\nu =180^\circ$  at $\rho_N=\rho_0$, so that the
asymmetry is slightly  smaller than for  $S/A= 1$.
At  higher density the asymmetry also becomes smaller, 
particularly when $S/A=2$. 

In any condition the neutrino production becomes larger in 
a direction parallel to the magnetic field ${\vB}$,
and smaller in the opposite direction.
The net result is that  the magnetic field increases the momentum flux 
of neutrinos emitted along the north magnetic polar direction 
while  decreasing the flux in the south polar direction. 
This magnetic contribution effect on the production process turns out 
to be of the same sign and magnitude as  the absorption process. 
Hence, the total asymmetry induced by the magnetic field from both processes 
should be about twice that from absorption alone \cite{MYCHKMR12}.

In summary,
we have calculated the magnetic contribution to the neutrino production 
through the direct URCA process and the absorption during transport.  
We have utilized an isoentropic model for
the proto-neutron star and employed RMF theory (with and without $\Lambda$ particles) for the EOS and to compute the  production cross-section.
The asymmetry in the absorption becomes larger at $\rho_B=\rho_0$
than that in our previous calculation based upon an isothermal model.
Furthermore, the asymmetry in the production cross-section is found to
be  also enhanced by the magnetic-field with the same magnitude and sign
as in the absorption process.

Since the scattering process also enhances
the neutrino asymmetry \cite{MYCHKMR12}, we can conclude that the magnetic-field effect causes 
asymmetric neutrino emission from a PNS through  the combination of the  production process as well as the absorption and the scattering. 
Therefore, the neutrino emission asymmetry from the neutrino sphere should be significantly  larger than previously estimated.
In future work we will consider  all magnetic effects
from the above three processes in a calculation of  pulsar kick velocities 
\cite{MYCHKMR12} and  spin-down \cite{MHKYKCRM13}.

\bigskip
This work was supported in part by the Grants-in-Aid for the Scientific
Research from the Ministry of Education, Science and Culture of
Japan~(26105517, 24340060, 21105512, 21540412)
and also by  the National Research Foundation of Korea (2012M7A1A2055605).
Work at the University of Notre Dame (GJM) supported by the U.S. Department of Energy under Nuclear Theory Grant DE-FG02-95-ER40934. 

%

%\newpage

\end{document}